%Paper: cond-mat/9207024
%From: Per Hyldgaard <hyldgaar@pacific.mps.ohio-state.edu>
%Date: Mon, 20 Jul 92 13:04:03 EDT

%%%
%%% The paper has 10 figures (1,2,3,4a,4b,5a,5b,6,7a,7b).
%%% The postscript files for these figures are located after
%%% the `\end' line of the texfile.
%%%
%%% The beginning of each postscript file is identified by
%%% a `header', consisting of 3 lines naming the figure. To locate
%%% the headers, search for `*'.
%%%
\input jnl
\input eqnorder
\input reforder

\def\ss#1{{\scriptscriptstyle #1}}
\def\pmb#1{\setbox0=\hbox{#1}
           \kern-.025em\copy0\kern-\wd0
           \kern.05em\copy0\kern-\wd0
           \kern-.025em\raise.0433em\box0}
\def\simgt{\mathrel{\lower .3ex \rlap{$\sim$}\raise .5ex \hbox{$>$}}}
\def\simlt{\mathrel{\lower .3ex \rlap{$\sim$}\raise .5ex \hbox{$<$}}}
\def\undertext#1{$\underline{\smash{\hbox{#1}}}$}
\def\tr#1{\hbox{Tr}\{ #1\} }
\def\la{\langle}
\def\ra{\rangle}
\def\ep{\epsilon}
\def\w{\omega}
\def\dt#1{{{d#1}\over{dt}}}
\def\sm#1{\ss{-#1}}
\def\psm{\phantom{-}}
\def\Int{\hbox{Int}}
\def\fancyd{\delta_{\ss{A,B=\left\{{ L\atop R }\right\} }}}
\def\ec{E_\ss{C}}
   \def\Cl{C_\ss{L}}
   \def\Cr{C_\ss{R}}
\def\Rl{R_\ss{L}}
   \def\Rr{R_\ss{R}}
\def\Vl{V_\ss{L}}
   \def\Vr{V_\ss{R}}
   \def\Vm{V_\ss{M}}
   \def\Vp{V_\ss{p}}
\def\kb{k_\ss{B}}
\def\ga#1{\gamma \left(#1\right)}
   \def\gl#1#2{\Gamma ^\ss{L}_{#1\rightarrow #2}}
   \def\gr#1#2{\Gamma ^\ss{R}_{#1\rightarrow #2}}
   \def\gls#1#2{\Gamma ^\ss{L}_\ss{#1\rightarrow #2}}
   \def\grs#1#2{\Gamma ^\ss{R}_\ss{#1\rightarrow #2}}
   \def\Gl{\Gamma ^\ss{L}}
   \def\Gr{\Gamma ^\ss{R}}
   \def\dgl#1#2{\delta\Gamma ^\ss{L}_\ss{#1\rightarrow #2}}
   \def\dgr#1#2{\delta\Gamma ^\ss{R}_\ss{#1\rightarrow #2}}
   \def\g#1#2{\Gamma _{#1\rightarrow #2}}
   \def\gs#1#2{\Gamma _\ss{#1\rightarrow #2}}
   \def\dg#1#2{\delta\Gamma _\ss{#1\rightarrow #2}}
\def\br{{\pmb {$\rho$}} (t)}
   \def\brz{\pmb {$\rho$}^\ss{\bf (0)}}
   \def\rb{\pmb {$\bar \rho$} ^\ss{\bf(0)}}
   \def\ro#1{\rho _{#1}^\ss{(0)}}
\def\bm{{\bf M}}
   \def\M#1#2{M_{#1#2}}
   \def\mb{{\overline {\bf M}}}
   \def\mbar{{\overline {M}}}
   \def\bdm{\pmb{$\delta $}{\bf M}}
\def\bn{{\bf N}}
   \def\var{\hbox{var}(n)}
   \def\Nl{N_\ss{L}}
   \def\Nr{N_\ss{R}}
   \def\Nmin{N_{min}}
   \def\Nmax{N_{max}}
\def\ir{I_\ss{R}}
   \def\il{I_\ss{L}}
   \def\ia{I_\ss{A}}
   \def\ib{I_\ss{B}}
\def\bvl{{\bf v^\ss{L}}}
   \def\bvr{{\bf v^\ss{R}}}
   \def\bva{{\bf v^\ss{A}}}
   \def\bvb{{\bf v^\ss{B}}}
   \def\bv{{\bf v}}
   \def\vr#1#2{v^\ss{R}_{i,j}}
   \def\vl#1#2{v^\ss{L}_{i,j}}
   \def\vab{{\bf \bar v^\ss{A}}}
   \def\vbb{{\bf \bar v^\ss{B}}}
   \def\vlb{{\bf \bar v^\ss{L}}}
   \def\vrb{{\bf \bar v^\ss{R}}}
\def\bpt{{\bf P}(t)}
   \def\P#1#2{P_{#1#2}}
   \def\bptm{{\bf P}(-t)}
   \def\ptb{{\bf \bar P(t)}}
   \def\ptmb{{\bf \bar P(-t)}}
   \def\dP#1#2{\delta P_{#1#2}}
   \def\bdp{\pmb{$\delta$}{\bf P}}
\def\lam{\lambda}
   \def\xl{{\bf x _\lambda }}
\def\dv#1{\frac{\partial #1}{\partial V}}
\def\dvx#1{\frac{\partial #1}{\partial \Vp}}

\title{Zero Frequency Current Noise for
the Double Tunnel Junction Coulomb Blockade}
\author{Selman Hershfield, John H. Davies,$^{\dag}$
Per Hyldgaard,
Christopher J. Stanton,$^{\ddag}$
and John W. Wilkins}
\affil{Department of Physics, 174 West 18th Ave., The
Ohio State University, Columbus, OH 43210}

\abstract{We compute the zero frequency current noise
numerically and in several limits analytically
for the coulomb blockade problem consisting of
two tunnel junctions connected in series.
At low temperatures
over a wide range of voltages, capacitances, and resistances
it is shown
that the noise measures the variance in the
number of electrons in the region between the two tunnel junctions.
The average current, on the other hand, only measures the mean
number of electrons.  Thus, the noise provides additional information
about transport in these devices which is not available from
measuring the current alone. }

\leftline{1992 PACS 72.70.+m , 73.40.Gk \hfil}
\endtopmatter

\head{\bf I. Introduction}
When the charging energy of a tunnel junction is larger than the
temperature, electron tunneling events across the junction become
correlated.
These correlations lead to a variety of phenomena which fall under
the rubric of single electron charging effects and the
Coulomb blockade.\refto{oldCB1,oldCB2,originalcalc,Likharev}
Recently, there has been renewed interest in the  Coulomb blockade
because of the wealth of new physical realizations:
metal-insulator-metal tunnel junctions with small metal
particles in the insulator,\refto{manyStair,fewStair,oneStair}
lithographically patterned tunnel junctions,\refto{Fulton}
scanning tunneling microscopy of small metal droplets,\refto{STM}
narrow insulating wires,\refto{Webb}
and even thin crossed wires.\refto{Gregory}
In some cases
the technology has advanced to such an extent that one
can consider making practical devices based on the coulomb blockade.
\refto{Mooij,Likharev}

The theoretical work in this area
has focused on the average current in either DC or
AC measurements.
There has been some work examining the effect of
noise in the external circuit on the average current;\refto{Clarke}
however,
there have not been any studies until recently treating the current noise as
an interesting phenomena in itself.\refto{noiseexp,ournoisecalc,noisecalc}
In this paper we compute the
zero frequency current noise in one of the simplest coulomb blockade
devices consisting of two tunnel junctions connected in series.
We apply the same master equation used to compute the average
current.\refto{originalcalc,numericalcalc,analyticcalc}
This work is a direct outgrowth of an earlier
paper in which we computed the noise with a similar equation, which
did not have charging effects explicitly.\refto{earlier}

We study the zero frequency as opposed to the finite frequency noise
because the time scale at which the frequency dependent
noise begins to show structure
is the ``RC" charging time of the system, which is quite small
($\sim  10^{-8}$ sec).\refto{STM})
Noise experiments, on the other hand, are typically done in the
regime below $10^{5}$ sec$^{-1}$.
Also, because this calculation is based on the usual master equation
for the coulomb blockade, it will contain thermal noise and shot noise,
but not 1/f noise, which is often seen in tunnel junctions at low
frequencies.\refto{Rogers}
1/f noise in tunnel junctions is usually assumed to be due to defects
in or near the junction.
These defects are not contained in our
model.  Although 1/f noise will dominate at the lowest frequencies,
there is often an intermediate regime between the very low frequency
regime and the high frequency regime where shot noise dominates.\refto{Tsui}
By focusing on the thermal and shot noise as opposed to the 1/f noise
we are able to make quantitative predictions for the noise.

The zero frequency current noise is interesting for several
reasons.  First, it is a measurable effect.  It may limit
the accuracy of some average current measurements.  Secondly, we shall
see that for this problem the noise provides a measure of the variance
in the charge fluctuations as a function of time.  The average current
only provides information about the average charge between the two
capacitors.  Finally, the noise versus current curve has
a rich structure, which is particularly revealing
when rescaled by the average current.
This can be used either to fit the parameters in the model
more accurately or to provide a consistency check for parameters
already determined from the average current.   Thus, noise measurements
provide an additional test for the underlying rate equation used to
describe transport in this simple coulomb blockade device.

This paper is organized into three parts: formalism (Sec. II.),
analytic results (Sec. III.), and numerical results (Sec. IV.).
In the formalism section the model and
underlying master equation are reviewed, the formulas for the
average current and noise are introduced, and the numerical technique used to
compute the current and noise is discussed.
Next in the analytic results section
we calculate the noise in four limits:
(i) the zero voltage limit, (ii) the region where the
voltage is larger than the temperature but still smaller than the
charging energy, (iii) the region just after the onset of current flow,
and (iv) the large voltage limit.
The details of the fluctuation dissipation theorem (zero voltage limit)
for this system are presented in the appendix.
Finally, the noise is calculated numerically and compared
to the analytic results.
All the results are summarized in Sec. V.

\head{\bf II. Formalism}
\subhead{\bf A. System}
The system which we study consists of two tunnel junctions connected
in series (see Fig. 1(a)).
The junctions are denoted by $L$ for left and $R$ for right, and
have resistances $\Rl$ and $\Rr$ and capacitances $\Cl$ and $\Cr$,
respectively.  We also introduce the net
resistance and capacitance of the junctions:
$R=\Rl + \Rr$ and $C=\Cl + \Cr$.
The capacitance $C$ is the capacitance which enters the
charging energy of the junctions, $\ec = e^2/(2C)$,
even though the capacitance of the two capacitors in series
is $(\Cl ^{-1}+\Cr ^{-1})^{-1}$.
To drive the current
a voltage $V$ is applied to the left junction, while the
voltage on the right junction is set to zero.
The voltage in the central or middle region between the two junctions, $\Vm$,
fluctuates depending on
the number of excess electrons in this region, $n$.
The voltage drops across the left
junction, $\Vl - \Vm (n)$, and the right junction, $\Vm (n) - \Vr$, are
found using classical electrostatics to be
$$
\eqalignno{
\Vl  -\Vm (n) &= \frac{\Cr}C V + \frac{ne}C + \Vp &(1.1)
\cr
\Vm (n) - \Vr  &= \frac{\Cl}C V - \frac{ne}C -\Vp .&(1.2)
\cr
}
$$
Here and in all following discussions the magnitude of
an electron's charge is $e$ so that $-e$ is the electron's charge.
The additional voltage $\Vp$ has been included to account for any
misalignment of the Fermi level in the middle region with the
Fermi levels of the left and right leads when $V$ and $n$ are zero.
\refto{oldCB2,analyticcalc}
One can also include an external gate on the central
region,\refto{analyticcalc}
which has a similar effect to $\Vp$.
For simplicity we omit such a gate.

\subhead{\bf B. Master Equation}

The transport through the tunnel junctions is governed by four tunneling
rates: the rate for electrons to tunnel onto the central region
from the left $(\Gamma ^\ss{L}_{n\rightarrow n+1})$ and right
$(\Gamma ^\ss{R}_{n\rightarrow n+1})$ and the rate for electrons to
tunnel off of the central region to the left
$(\Gamma ^\ss{L}_{n\rightarrow n-1})$ and right
$(\Gamma ^\ss{R}_{n\rightarrow
n-1})$.\refto{originalcalc,numericalcalc,analyticcalc}
The number of excess electrons in the central region between
the two junctions is $n$.
These rates are computed via Fermi's golden rule.
In order to write the rates we introduce a function $\ga \ep$:
$$
\ga \ep = \frac \ep{1-e^{-\beta \ep}} .
\eqno (1.3)
$$
For energies $\ep$ larger than the temperature, $\kb T$, $\ga \ep$
is approximately $\ep$, showing that the tunneling rate increases
as the allowed phase space is increased.
For energies $\ep$ less than $-\kb T$, $\ga \ep$ is exponentially
suppressed: $\ga \ep \approx |\ep |e^{-\beta |\ep |}$, since the energy
for these processes must come from thermal fluctuations.
In terms of
the charging energy, $\ec =e^2/(2C)$, the tunneling rates
are\refto{originalcalc,numericalcalc,analyticcalc}
$$
\Gamma ^\ss{L(R)}_{n\rightarrow n\pm 1} =
\frac {1}{e^2R_\ss{L(R)}}
\,\ga {\pm e[\Vm (n) - V_\ss{L(R)} ] - \ec} .
\eqno (1.5)
$$
The energies which enter the rates of Eq. \(1.5) are the
voltage drops offset by the charging energy, indicating
that tunneling is suppressed for voltages smaller than the
charging energy.

The state of the system at time $t$ is described by the probability
$\rho _n(t)$ that there are $n$ electrons in the central region.
Clearly the sum of the $\rho _n(t) $ is unity:
$\sum _n \rho _n(t) = 1$.
The time evolution of $\rho _n(t)$ is governed by
the net rate, $\g ij=\gr ij +\gl ij$, to go from $i$ to $j$
excess electrons in the middle region.
By incorporating the rates
$\g ij$  in a matrix $\bm$,
$$
\M ij = \cases{\g ji,&if $i=j\pm 1$,\cr
              -\g j{j+1}-\g j{j-1}, &if $i=j$,\cr
                0,&otherwise, \cr}
\eqno (1.7)
$$
the Master equation for the
time evolution of $\br $ may be written succinctly as
$$
\frac{d\br}{dt} = \bm \br .\eqno(1.6)
$$
A direct consequence of Eqs. \(1.7) and \(1.6) is
that the sum of the $\rho _n(t) $ is independent of time because
$$
\sum _i \M ij =0 .\eqno(1.8)
$$
Equation \(1.8) in turn implies that the matrix $\bm$ has a
zero eigenvalue so there is a steady state solution,
i.e., a vector $\brz $ which satisfies
$$
0= \bm \brz .\eqno (1.9)
$$

\subhead{\bf C. Current}

Although only the net transition rate, $\g ij$, enters in computing
$\br $, the current depends on whether an electron travels to
the right or left.  For example, a transition from $n$ to
$n+1$ electrons in the central region has the rate $\g n{n+1}$.
If the additional electron tunneled from the right, this process
gives a negative contribution to the number current and a positive
contribution to the electrical current.  On the other hand, if the
additional electron tunneled from the left, there would be a positive
contribution to the number current and a negative contribution to
the electrical current.
To take into account the difference in sign for the two processes,
we introduce two new matrices, $\bvl$ and $\bvr$, which
contain the rates for tunneling across the left and
right junctions, respectively.
The sign of a matrix element is positive if a process gives
a positive contribution to the number current from left to right
and negative if a process gives a negative contribution to the number
current.
$$
\hbox{v}^\ss{L(R)}_{ij} = \cases{\ss{-(+)}\Gamma ^\ss{L(R)}_{j\rightarrow i},
                                                            &if $i=j-1$,\cr
               \ss{+(-)}\Gamma ^\ss{L(R)}_{j\rightarrow i},&if $i=j+1$\cr
                0,&otherwise.\cr }
\eqno(1.10)
$$
With these definitions the electrical current across the left $(\il )$
and right $(\ir )$ junctions at time $t$ are
$$
I_\ss{L(R)}(t) = -e \sum _i(\hbox{v}^\ss{L(R)} \br  )_i \equiv
-e\tr {\hbox{v}^\ss{L(R)} \br  } .\eqno (1.11)
$$
In Eq. \(1.11) we have defined a trace of a vector to be the sum
of its elements.
Both the currents $\il$ and $\ir$ are
positive when the electrical current goes from left to right.
To obtain the steady state current from Eq. \(1.11)
$\br $ is replaced by $\brz$.

In general the current in the left and right junctions will be
different; however, in the zero frequency limit they must be the same
because of the continuity equation.
To derive the continuity equation we introduce the number matrix
$\bn$, where $N_{ij} = i \times \delta _{ij}$.
The expectation value for the number of electrons in the central
region at time $t$ is
$$
N(t) = \tr{\bn \br} .\eqno (1.12)
$$
Using the rate equation for the probability, Eq. \(1.6), the time
derivative of $N(t)$ is
$$
\frac {dN(t)}{dt} =
\tr{\bn \bm \br} .\eqno (1.13)
$$
The key observation is that $\bvl -\bvr$ is the
commutator of $\bn$ and $\bm$.
$$
\bvl - \bvr = [\bn ,\bm ] .\eqno (1.14)
$$
Using Eqs. \(1.8) and (1.14) the time derivative of
the charge in the central region may be written as
$$
-e\frac {dN(t)}{dt} =
-e \tr{[\bn ,\bm ]\br} = \il (t) -\ir (t),
\eqno (1.15)
$$
which is the continuity equation.
As noted above in steady state, $dN(t)/dt=0$, and
$\il$ is equal to $\ir$.

\subhead{\bf D. Noise}

To define the noise we introduce the propagator $\bpt$ which gives
the time evolution of $\rho _n(t)$
$$\bpt = \exp(\bm t).
\eqno (1.16)
$$
The conditional probability that there are $m$ electrons in the middle region
given that there were $n$ electrons at $t=0$ is $P_{m,n}(t)$.
It is understood that $t$ is positive in Eq. \(1.16).
With $\bpt$
we can compute all possible correlation functions between the
density and the current.\refto{classicalinbook,LandNoise}
For example, the density-density correlation function is
$$
\eqalign{
\langle N(t) N(0) \rangle &= \theta (t) \tr{ \bn \bpt \bn \brz } \cr
&+\theta (-t) \tr{ \bn \bptm \bn \brz } .\cr
}\eqno (1.17)
$$
This equation has a simple physical interpretation.  Initially, the
probability distribution is the steady state solution, $\brz$.
The number of the electrons in the central region is measured by $\bn$
at this initial time. Next, the system is propagated forward in time
via ${\bf P}$, and the number is again measured with $\bn$.
For $t>0$ the initial time is $t=0$ and the final time is $t$, while
for $t<0$ the initial time is $t$ and the final time is $0$.
In a similar manner the density-current correlation function is
($A=R$ or $L$)
$$
\eqalign{
\langle N(t) \ia (0) \rangle =&-e\,\theta (t) \tr{ \bn \bpt \bva \brz } \cr
                             &-e\,\theta (-t) \tr{ \bva \bptm \bn \brz }  .\cr}
\eqno (1.18)
$$
The only difference between Eqs. \(1.18) and \(1.17) is that the
matrix $\bva$ measures the current across one of the junctions instead
of the number of electrons in the central region measured by $\bn$.

In analogy with Eqs. \(1.17) and \(1.18)
the current-current correlation function
should have two matrices $\bva$ and $\bvb$ for measuring
current at the two times: $0$ and $t$.
Although such a term includes the correlation between two different tunneling
events, it does not include the self-correlation of a given tunneling
event with itself.
In our earlier paper\refto{earlier}
we gave a formal derivation of the self-correlation term as well as
the correlation between different tunneling events.
Here we put in this self-correlation term by hand.
Suppose there is a current pulse across the right junction
between $0$ and $dt$.  The number current is $(dt)^{-1}$ during the
time interval $dt$, and the number
current squared is $(dt)^{-2}$.  The probability of this happening
$dt$ times the average number current to the right.
Thus, it would seem that the self-correlation term for
$\la \ir (t) \ir (0) \ra $ is just $eI\delta (t)$.
However, the current can be either positive or negative, and the
self-correlation term is always positive.  Thus, in computing the
self-correlation term we must take the absolute value of the
matrix elements of $\bvl$ and $\bvr$ used in computing the current.
These new matrices may be written succintly as
$([\bn , \bvl ])_{i,j} = |(\bvl )_{i,j}|$ and
$([\bvr , \bn ])_{i,j} = |(\bvr )_{i,j}|$.
Thus, the current-current correlation function is
($A,B = $ $R$ or $L$)
$$
\eqalign{
\la I_\ss{A}(t) I_\ss{B}(0) \ra &= e^2\theta (t)\tr{ \bva \bpt \bvb \brz } \cr
                       &+e^2\theta (-t)\tr{ \bvb \bptm \bva \brz }  \cr
            &\pm \fancyd e^2\delta (t)\,\tr{[\bn ,\bva ]\brz } . \cr}
\eqno (1.19)
$$
The first two lines of this equation contain the correlation
between two different tunneling events, while the last line is
the self-correlation term.

As for the average current, the continuity equation identity,
Eq. \(1.14), can be used to show that any
occurrence of $(\il - \ir)$ may be replaced by $-e(dN/dt)$.
$$
\eqalignno{
-e\frac {d}{dt}\la N(t) N(0) \ra &=
\la (\il (t)-\ir (t))N(0) \ra &(1.20) \cr
-e\frac {d}{dt}\la N(t) \ia (0) \ra &=
\la (\il (t)- \ir (t)) \ia (0) \ra . &(1.21) \cr
}$$
In the zero frequency limit this implies that both the current-current
correlation function and the current-density correlation functions
are independent of where the current is measured.
The noise, $S_\ss{AB}(\w )$, is now defined as the fourier
transform of the current-current correlation function,
$\la \ia (t) \ib (0) \ra$, with its long time behavior subtracted off.
$$
S_\ss{AB}(\w ) = 2\int^\infty_{-\infty} dt e^{i\w t}
\left( \la \ia (t) \ib (0) \rangle -\langle I\rangle ^2 \right) .\eqno (1.22)
$$
The continuity equation (Eq. \(1.21)) and the relation $S_\ss{AB}(\w )
=S_\ss{BA}(-\w )$ imply that
correlation functions for all possible choices of $A$, $B$ = $R$, $L$
are equal in the zero frequency limit, $\w = 0$.
As discussed in the introduction we will always be taking the zero
frequency limit because it is the easiest to study experimentally.

\subhead{\bf E. Numerical Technique}
In order to compute the noise
it is useful to symmetrize the matrix $\bm$.
The key observation is that this system obeys detailed
balance (see Eq. \(1.9)):
$$
\g ij\ro i = \g ji\ro j .\eqno (1.23)
$$
A direct consequence of Eq. \(1.23) is that $M_{ij} \ro j = M_{ji} \ro i$,
allowing us to define a symmetric matrix $\mb$,
$$
\mbar _{ij} \equiv
\sqrt{ \ro j /\ro i}\,\M ij =
\sqrt {\M ij \M ji } .
\eqno (1.24)
$$
In a similar manner we define $\vab$ as $(\ro j/\ro i)^{1/2}\,\bva$ and
$ {\bar \rho}_i^\ss{(0)}$ as the square root
of  ${\rho _i^\ss{(0)}}$
so that the expectation values
may also be written in a symmetric form, e.g.
$$
\ia = -e\rb \vab \rb .\eqno (1.25)
$$
The matrix $\mb$ is written in terms of its eigenvalues, $\lam \le 0$,
and eigenvectors, $x_\lam$, as
$$
{\mbar}_{ij} = \sum _\lam \lam (x_\lam )_i  (x_\lam )_j.
\eqno (1.26)
$$
Using this same representation, the noise, $S_\ss{AB}(\w )$,
is given by
$$
\eqalign{
S_\ss{AB}(\w ) =
&2e^2\sum _{\lam \ne 0} (\rb \vab \xl ) \frac {1}{-\lam -i\w}(\xl \vbb \rb )
\cr
+&2e^2\sum _{\lam \ne 0}(\rb \vbb \xl ) \frac {1}{-\lam +i\w}(\xl \vab \rb )
\cr
\pm&\fancyd 2e^2 \,(\rb [\bn ,\vab ] \rb)  . \cr}
\eqno (1.27)
$$
In this paper we truncate the matrix $\mb$ to include some finite
number of states and then diagonalize it to determine its eigenvalues
and eigenvectors.  The eigenvector with zero eigenvalue is $\rb$.
Equations \(1.25) and \(1.27) are used to compute the current and noise.

\head{\bf III. Limits}
Before discussing the results of our numerical evaluation of the zero
frequency noise, in this section we compute the noise in four limits
which may be treated analytically.  The cases discussed are:
A. the zero voltage limit, where the noise is related to the
conductance via the fluctuation dissipation theorem;
B. the thermally activated conduction regime,
where the voltage drop is large compared to the temperature
but still small compared to the charging energy;
C. the two state region, where the voltage is larger than the charging energy
but small enough that only two charge states are allowed; and
D. the large voltage bias limit.

\subhead{\bf A. Zero voltage}
Even though
we will focus on potential drops,
$eV$, large compared to the thermal energy, $\kb T$,
any technique for computing the noise must reproduce
the fluctuation dissipation theorem\refto{Johnson,Nyquist} at zero bias:
$$
S = 4\kb T \, G(V=0),
\eqno (FDT)
$$
where $G$ is the differential conductance, $G= \partial I/\partial V$.
In the appendix we show how Eq. \(FDT) follows from
the definitions of the rates (Eq. \(1.5))
and the expression for the noise (Eq. \(1.27)).
In Sec. IV. we also verify that our
numerical algorithm produces the fluctuation dissipation
theorem at zero bias.

\subhead{\bf B. Thermally activated conduction}
When the voltage is large compared to the temperature,
the thermal noise is small compared to the shot noise, which comes
from the current flowing.
Even with $e|V| \gg \kb T$, if the voltage is small compared to
charging energy, the conduction is thermally activated because
the voltage drop alone does not provide enough energy to overcome
the charging energy.  In this section we show that in this regime
$(\ec \gg e|V| \gg \kb T)$ the noise is related to the current
by the standard shot noise relation
$$
S=2e|I|. \eqno (standard)
$$

In this limit
the steady state probability distribution function
$\rho _n^\ss{0}$ is strongly
peaked about one value of $n$ except for special values of $\Vp$
where two states are equally occupied.
We shall assume that we are not at one of these special points and
call the state which is most likely occupied the $n=0$ state.
To describe the transport it is
sufficient to keep only three states: the $n=0$ state and the
two states immediately accessible from this state: $n=\pm 1$.
At $T=0$ the matrix $\bm$ thus becomes:
$$
\bm (T=0) = \pmatrix{
-\gs{-1}{0} & 0 &         0  \cr
\psm\gs{-1}{0} & 0 &  \psm\gs{1}{0} \cr
        0  & 0 & -\gs{1}{0}. \cr
}\eqno (2.18)
$$
The eigenvector of $\bm (T=0)$ with zero eigenvalue is $(0,1,0)$,
indicating that at zero temperature
only $\rho _{n=0}^\ss{0}$ is nonzero.
For a finite temperature there is a small correction
$\bdm \equiv \bm - \bm(T=0)$,
which allows transitions to the states $n=\pm 1$.
$$
\bdm = \pmatrix{
-\dg{-1}{0} &  \dg {0}{-1}             &          0  \cr
\psm\dg{-1}{0} & -\dg {0}{1} -\dg {0}{-1} &  \psm\dg {1}{0} \cr
         0  &  \dg {0}{1}              & -\dg {1}{0} \cr
}\eqno (2.19)
$$
There are analogous zero temperature and finite-temperature
correction terms for $\bvl$ and $\bvr$.

Because we have reduced $\bm$ to a $3\times3$ matrix,
equations \(1.25) and \(1.27) for the current and noise can
easily be expanded to linear order in the $\delta \Gamma$'s:
$$
\eqalignno{
\frac Ie &=
    \left[ \frac{\gls 10}{\gs 10}\,\dgr 01
          +\frac{\grs {-1}0}{\gs {-1}0}\,\dgl 0{-1} \right]
  - \left[ \frac{\grs 10}{\gs 10}\,\dgl 01
          +\frac{\gls {-1}0}{\gs {-1}0}\,\dgr 0{-1} \right]
&(2.20) \cr
\frac S{2e^2}&=
       \left[ \frac{\gls 10}{\gs 10}\,\dgr 01
          +\frac{\grs {-1}0}{\gs {-1}0}\,\dgl 0{-1} \right]
  +    \left[ \frac{\grs 10}{\gs 10}\,\dgl 01
          +\frac{\gls {-1}0}{\gs {-1}0}\,\dgr 0{-1} \right] .
& (2.21) \cr}
$$
These equations have a simple physical interpretation.
Conduction in this thermally activated
regime follows from the occurrence of
a rare transition from
$n=0$ to $n=\pm1$, followed
by a rapid decay back into the $n=0$ state.
Thus, for example, there could be a thermal fluctuation causing
an electron to tunnel from the right lead to the central region
$(\dgr 01)$.
This thermal fluctuation is followed shortly by the electron
tunneling off to the right lead $(\grs 10)$ or to the left lead $(\gls 10)$.
If an electron tunnels to the right lead, then there is no contribution
to the current, while if the electron tunnels to the left lead
there is a positive contribution to the electrical current.
Thus, only a fraction of the time, $\gls 10/\gs 10$,
does this thermally activated
process contribute to the current.  This explains the first term
in Eq. \(2.20): $(\gls 10/\gs 10)\dgr 01$.
The other terms in this equation have similar interpretations.

The noise in this regime only contains the self-correlation
term because we have independent thermally activated
processes which are separated by long periods in time.
Thus, the only difference between the current and the noise
is one sign change resulting from some processes giving a negative
contribution to the current and a positive self-correlation
term for the noise.  Dividing Eq. \(2.21) by Eq. \(2.20),
it would seem that their ratio is not 2e;
however, because the voltage is large compared to the temperature,
two of the activated
rates are always much larger than the other two.
For $I>0$, $\dgr 01$ and $\dgl 0{-1}$ are much larger than
$\dgl 01$ and $\dgr 0{-1}$ by a factor exponential in $e|V|/\kb T$,
while for $I<0$ the situation is reversed.
Therefore,
the noise is indeed equal to 2e times absolute value of the current
(Eq. \(standard)).

\subhead{\bf C. Two state regime}
Eventually the voltage becomes larger than the charging energy,
and there is current flow even at zero temperature.
As discussed above,
in the thermally activated regime only one charge
state is allowed at zero temperature.
At the onset of non-thermally activated current flow, there
are two allowed charge states at $T=0$.
It is a good approximation to keep only these two states for
some range in temperature, $\kb T \ll \ec$.
Our two-state approximation is similar but not identical
to the one used in Ref.~\cite{twostate}.
They compute $\brz$ using only two states, and then allow transitions
to a third state in computing the current (Eq. \(1.25)).
We do not allow transitions to a third state; however,
in the regime of interest here, where only two states are
energetically accessible, the two approaches give the same
result for the current.

Since only two states are involved, there are only two rates.
In the case $I>0$ one rate, $\Gr$, increases the number of electrons
by tunneling across the right junction, while the other rate, $\Gl$,
decreases the number of electrons by tunneling across the left junction.
These rates depend on the voltage drop across the sample and
can be determined from Eqs. \(1.3) and \(1.5).  In the next section
we will also give the limiting form of these rates at zero temperature.
Since $\bm$ in this limit is a $2\times2$ matrix,
$$
\bm = \pmatrix{
-\Gr & \Gl \cr
\Gr & -\Gl \cr },
\eqno (2stM)
$$
it is simple to solve for the current and the noise:
$$ \eqalignno{
I &= e \,\, \frac{\Gl\Gr}{\Gl +\Gr} \Big|_V &(2stA) \cr
S &= 2eI \,\, \frac{(\Gr )^2+(\Gl )^2}{(\Gr +\Gl )^2} \Big|_V. &(2stB)\cr
}$$
We wish to emphasize here that although Eqs. \(2stA) and \(2stB)
look simple, they contain a great deal of structure because
the rates $\Gl$ and $\Gr$ depend on voltage.
This will be shown explicitly in Sec. IV. when we plot this
result along with the numerical results for finite temperatures.
{}From Eqs. \(2stA) and \(2stB) we
can see that $I$ is largest when two rates are equal.
Also, the noise is suppressed from its uncorrelated value,
$S=2e|I|$.
This suppression is to be expected because having only
two allowed states in the central region introduces
correlations between tunneling events.
If the system is at one of the allowed states, say the state
with more electrons, then another electron can not tunnel onto
the central region until an electron tunnels off.
This kind of correlation did not appear in the thermally activated
regime because an electron tunnels off so quickly after tunneling on
that it has no effect on subsequent tunneling events.

\subhead{\bf D. High voltage limit}
We have now computed the noise for $V=0$, $\ec \gg e|V| \gg \kb T$,
and $e|V| \simgt \ec$.  In this section we compute the
noise for $e|V| \gg \ec$, where we expect the current-voltage
characteristic to be almost linear.  Even though we compute the
noise in the asymptotic high voltage limit, in Sec. IV. we will see
some of the approximations derived here work down to $e|V| \sim \ec$.

Because we are interested in the limit where the voltage is larger
than the other energy scales, $\kb T$ and $\ec$, we set the temperature
equal to zero.
This means that $\ga \ep$ in Eq. \(1.3)
takes on a very simple form:
$\ga \ep = \ep \,\theta (\ep )$.
If the sample is biased so that the electrical current flows from
left to right, then two kinds of processes are important at $T=0$:
tunneling across the right junction to the central region
$(\gr n{n+1} )$
and tunneling from the central region to the left lead $(\gl n{n-1} )$.
The rates for these processes are
$$
\eqalignno{
\gl n{n-1} &= \frac 1{\Rl C}\, (n-\Nl )\theta (n-\Nl ) &(2.22) \cr
\gr n{n+1} &= \frac 1{\Rr C}\, (\Nr -n)\theta (\Nr -n) , &(2.23) \cr
}$$
where the numbers $\Nl$ and $\Nr$ are given by
$$
\eqalignno{
\Nl &= -\frac{\Cr V}{e} - \frac{C\Vp}{e} + \frac{1}{2} &(2.24) \cr
\Nr &= \phantom{-}\frac{\Cl V}{e} - \frac{C\Vp}{e} - \frac{1}{2}.&(2.25) \cr
}$$
These rates are illustrated in Fig. 1.(b).
The minimum allowed state, $N_{min}$,
is the largest value of $n$ for which $(n-\Nl )<0$.  Similarly,
the maximum allowed state, $N_{max}$, is the smallest $n$
for which $(\Nr -n)<0$.  Assuming that $\Nl <0$ and $\Nr >0$, this means
that the values of $n$ with $\rho _n^\ss{0} \ne 0$ are
$$
N_{min}=
-\Int (-\Nl -1)  \le n \le \Int (\Nr +1)=N_{max} .
\eqno (allowed)
$$

Since $\ir$ and $\il$ are just the expectation values of $\gr n{n+1}$ and
$\gl n{n-1}$ respectively, it is tempting to call Eqs. \(2.22) and \(2.23) with
$n \rightarrow \la n \ra$ the current.  This is not correct because
$\gl n{n-1} = (n-\Nl )/(\Rl C)$
is only valid for $n \ge N_{min}+1$. For $n=N_{min}$
the rate $\gl n{n-1}$ is zero.
Similarly,
$\gr n{n+1} = (\Nr -n)/(\Rr C) $
is only valid for $n \le N_{max}-1$.
However, in the limit $\Rl \ll \Rr$ the rate for tunneling across the left
junction, $\gl n{n-1}$,
is much larger than the rate for tunneling across the right
junction as illustrated in Fig. 1(b).
This means that $\la n\ra$ is close to $N_{min}$.
Because $\rho ^\ss{(0)}_{N_{max}} \approx 0$, it is then a
good approximation to replace $n$ by $\la n\ra$ in Eq. \(2.23):
$$
I \approx e \frac 1{\Rr C}(\Nr - \la n \ra ) .
\eqno (meanl)
$$
In similar manner for $\Rr \ll \Rl$ the probability of being in
the state $n=N_{min}$ is small and we can approximate the current
by Eq. \(2.22) with $n\rightarrow \la n\ra$:
$$
I \approx e \frac 1{\Rl C}(\la n \ra -\Nl ) .
\eqno (meanr)
$$
Equations \(meanl) and \(meanr) show that the current
measures the average number of electrons in the middle region.

Up to this point we have not made any assumptions about the
voltage being much larger than the charging energy.  Now we
assume that $\Nl$ and $\Nr$ are integers.
This can only be true for a limited set of voltages;
however, in the large voltage limit
the current and noise should insensitive to whether
$\Nl$ and $\Nr$ are integers.  We verify this numerically in
the Sec. IV. by showing that the results predicted here are
asymptotically true in the high voltage limit.

If $\Nl$ and $\Nr$ are integers, then we have shown in
our earlier paper\refto{earlier} that the current and noise can be computed
exactly.  Here, we review the results
in the context of this problem.
For integer $\Nl$ and $\Nr$
the number electrons in the
middle region, $N(t)$, satisfies a simple rate equation
$$
\dt {N(t)} = \frac {\Nr -N(t)}{\Rr C} -
           \frac {N(t)- \Nl }{\Rl C} .
\eqno (2.26)
$$
This allows one to compute both the steady state current
$$
I = \frac e{RC}
\left( \frac{CV}e -1 \right) ,
\eqno (2.27)
$$
and to relate the zero-frequency noise to the
variance in the number of electrons in the
middle region, $\var$,
$$
S=2eI -  \frac {4e^2}{RC} \,\var
\eqno (2.28)
$$
The variance of $n$ is given by
$$
\var \,= \,\frac{\Rl ^{\sm 1}\Rr ^{\sm 1}}{(\Rl ^{\sm 1} +\Rr ^{\sm 1})^2}
\left( \frac{CV}e -1 \right)
\eqno (2.29)
$$
so the noise in the high voltage limit is
$$
S = 2eI \,\frac{\Rl ^{\sm 2} + \Rr ^{\sm 2}}{(\Rl ^{\sm 1} +\Rr ^{\sm 1})^2} .
\eqno (2.30)
$$
Note the formal similarity of these results with those of
the two state regime (Eqs. \(2stA) and \(2stB)).
An important difference between
these two sets of results is that $\Rl^\ss{-1}$ and $\Rr\ss{-1}$ are constant,
while $\Gl$ and $\Gr$ depend on voltage.
Thus, here $S/2eI$ goes to a constant asymptotic value, while before
$S/2eI$ depended on voltage.
In the next section we will show numerically that Eq. \(2.30)
is indeed valid in the high voltage limit, and Eq. \(2.28) is a good
approximation for much smaller voltages.

\head{\bf IV. Numerical Results}
In this section we present the results for computing the noise
and current via Eqs. \(1.25) and \(1.27).
There are six parameters in the model
besides the voltage: the resistances, $\Rl$ and $\Rr$,
the capacitances, $\Cl$ and $\Cr$,
the temperature, $T$, and the offset voltage, $\Vp$.
Two of these parameters can be eliminated by
rescaling the rates by $R=\Rr + \Rl$ and the energies by
the charging energy, $\ec = e^2/2C$.  Thus, our parameter space consists
of the resistance ratio, $\Rl/R$, the capacitance ratio, $\Cl /C$,
the temperature divided by the charging energy, $\kb T/\ec$,
and $e\Vp/\ec$.
In evaluating Eqs. \(1.25) and \(1.27) we will always truncate
the matrix $\bm$ because the higher charge states, $|n| \gg 1$,
are energetically forbidden, especially at low
temperatures and voltages.  In practice we have found
it is sufficient to keep the 15 states surrounding $n=0$.
With this many states
the steady state solution, $\rho ^\ss{(0)}_n$, is zero to within
our numerical accuracy for the highest $(n=7)$ and lowest states
$(n=-7)$.
Increasing the number of states does not change $\rho ^\ss{(0)}_n$.

\subhead{\bf A. Temperature Dependence}
For most of this section we will consider temperatures
far below the charging energy because this is where
charging effects are most important.
To start, however, we set the voltage equal to zero
and verify that our numerical algorithm satisfies the
fluctuation dissipation theorem (Eq. \(FDT)).
In Fig. 2 we have plotted both the noise, $S$, and the differential
conductance, $G=dI/dV$, as a function of temperature
for a pair junctions with $\Rl/R=0.01$ and $\Cl/C=0.01$.
As expected in a coulomb blockade problem,
the conductance and noise are suppressed for temperatures
much less than the charging energy
($\simlt 0.2 \ec$, here).   For temperatures greater than or of order
the charging energy, the conductance rises linearly with
temperature, indicating that the number of accessible states
is increasing with temperature.
In this same regime the noise approaches a constant value
as required by the fluctuation dissipation theorem.
We have divided the two curves to show explicitly that the fluctuation
dissipation theorem is satisfied.

\subhead{\bf B. Dependence on $\Vp$}
In Fig. 3 the current and noise versus voltage curves are
plotted for $C\Vp = 0$ and $C\Vp =0.25$.
The temperature is much less than the
charging energy, $\kb T= 0.01\ec$.
As expected for an asymmetric pair of junctions with
$\Rl = 0.01R$ and $\Cl = 0.01C$, the
curves exhibit the step-like structure of the coulomb staircase.
The primary effect of the offset $\Vp$ is to shift the I-V and S-V
curves.  To reduce the parameter space in subsequent plots
we will specialize to $\Vp =0$, where the I-V and S-V characteristics
are symmetric in the voltage.
In this figure
the fact that the noise and current curves are almost the same indicates
that the noise is close to its uncorrelated value:  $S=2eI$.
Careful examination shows that the noise is actually less than $2eI$.
The amount by which it is less than $2eI$ changes as a function
of voltage.  Thus, in subsequent plots we will also look at
the noise ratio, $S/2eI$, to bring out this structure.

\subhead{\bf C. Dependence on the capacitance and resistance ratios}
Restricting ourselves to $\kb T \ll \ec$ and $\Vp =0$,
the parameter space becomes two dimensional:
$(\Rl /\Rr ,\Cl /\Cr )$ or $(\Rl /R,\Cl /C )$.
It is well known that for asymmetric junctions with
$\Rl \ll \Rr$ and $\Cl \ll \Cr$ (or
$\Rl \gg \Rr$ and $\Cl \gg \Cr$)
the coulomb staircase structure as in Fig. 3 is seen,
while for asymmetric junctions with
$\Rl \ll \Rr$ and $\Cl \gg \Cr$ (or
$\Rl \gg \Rr$ and $\Cl \ll \Cr$)
no step-like structure is seen.
There is only a suppression of the current for small voltages.

These two limits can be understood as follows.
The resistances, $\Rl$ and $\Rr$, determine the overall scale for the
transition rates given in Eq. \(1.5).  If, for example, $\Rl \ll \Rr$, then
the tunneling rate across the left junction is much larger than the
tunneling rate across the right junction.
For a positive bias, $V>0$, electrons will tunnel off the middle
region to the left lead.  Thus, in the case $\Rl \ll \Rr$, the number of
electrons in the central region will be close to the minimum allowed number,
$N_{min}$ of Eq. \(allowed).
Using the approximate expression for the current when
$\Rl \ll \Rr$, Eq. \(meanl), there are steps in the I-V characteristic
when $\la n \ra$ jumps,
but no steps when $\la n \ra$ is approximately constant.

This raises the question: when does $\la n \ra$
jump discretely with increasing voltage?
While the resistances determine the overall scale of the rates,
the capacitances determine the energetics.  If $\Cl \ll \Cr$, then the
minimum allowed charge state at low temperatures is
$N_{min} =-\Int (\Cr V/e+1/2)$ (see Eqs. \(2.24) and \(allowed)).
On the other hand if $\Cl \gg \Cr$ the minimum
allowed charge state at low temperatures is $\Nmin =0$.  Thus, in the first
case the minimum allowed charge state changes with voltage while in
the second case it does not.
This means that for $\Cl \ll \Cr$ and $\Rl \ll \Rr$
the charge state will increase discretely as one increases
the voltage, while for $\Cl \gg \Cr$ and $\Rl \ll \Rr$
the charge state will remain approximately constant.

We have only discussed the two extreme limits when there are well
defined steps in the I-V characteristic and when there are no
steps in the I-V characteristic.  Clearly, one can go continuously
between these two cases by varying the resistance and capacitance
ratios.  In Fig. 4 we have computed the current and noise ratio, $S/eI$,
as a function of voltage for $\Cl /C= 0.01$ varying $\Rl /R$,
and in Fig. 5 we show
the same quantities for $\Rl /R =0.01$
varying $\Cl /C$.  For both $\kb T= 0.1\ec$ and $\kb T= 0.01\ec$ the
I-V characteristic does indeed go between a step like structure
for $\Rl \ll \Rr$ and $\Cl \ll \Cr$ and an offset only for
$\Rl \ll \Rr$ and $\Cl \gg \Cr$, or
$\Rl \gg \Rr$ and $\Cl \ll \Cr$.
In comparing Figs. 4 and 5 it is important to note that
although the dimensionless ratio in the abscissa
is the same in the two figures,
the range of voltages is different.
In Fig. 4 $\Cr$ is equal to $0.99C$, while $\Cr$ varies from
$0.99C$ to $0.05C$ in Fig. 5.
Thus, the curves at the top of Figs. 4 and 5 appear different,
while they are in fact almost the same when plotted in
terms of the unscaled voltage.

Another important difference between Figs. 4 and 5 is
the way one goes between the two extremes.
When varying the resistances (Fig. 4),
the steps become smoother, while in varying the capacitances (Fig. 5),
the steps remain sharp, but the region between them gets a finite slope.
In the first case when the resistances are changed, the probability
distribution function becomes more spread out in $n$ as $\Rl$ becomes
comparable to $\Rr$.  Thus, instead of having the distribution function
peaked near $N_{min}$ one obtains a much broader distribution function.
Consequently, as new charge
states become allowed, the distribution gradually shifts to lower $n$
instead of jumping immediately to the new $N_{min}$.
In the second case when the capacitances are changed, the distribution
remains fixed near the lower charge state, but the $\la n \ra$
in the current, $I \approx (\Nr - \la n \ra )/\Rr C$,
becomes less important relative to the $\Nr$ term.
In this case the jumps remain sharp, but become smaller in magnitude.

Next we turn to the noise ratio $S/2eI$ vs. current curves, which are the
new contribution of this paper.  The first observation to make is that
there is a rich variety of structure in the noise when rescaled by the
current.  This structure is not contained in the current.
For example, one might say that the $S/2eI$ curves in Fig. 5 are simply
related to the derivative of the current; however, there is no
simple way to take the derivative of the current curves in Fig. 4
and obtain the $S/2eI$ curves in that figure.

Using the structure in $S/2eI$, one can determine the parameters in the model
more accurately than one could with the current alone.
Alternately,
one can use the parameters obtained from the current to perform a
consistency check.  We illustrate this point by considering three
examples from Figs. 4 and 5.   In Fig. 4(b), the top three cases,
$R_\ss{L}/R =0.75$, 0.9, and 0.99,
have almost identical current-voltage characteristics, yet
their noise ratio curves are quite distinguishable.
As a second example, the steps in the $R_\ss{L}/R=0.5$ I--V curve are only
slightly more pronounced than the steps in the $R_\ss{L}/R=0.75$ curve;
however, the noise-ratio curves exhibit a change in curvature.
The upper curve has a downward cusp, while the lower curve
has an upward cusp.  Finally, we note that the $R_\ss{L}/R=0.1$
curve in Fig. 4(b) and the
$C_\ss{L}/(C_\ss{L}+C_\ss{R})=0.25$ or 0.5 curves
in Fig. 5(b) are quite similar,
although their noise ratio curves are very different.

Comparing Fig. 4(a) to 4(b) and Fig. 5(a) to 5(b),
there is considerable sharpening in the noise ratio curves as
one goes down in temperature.  The spiky structure of the
curves in Fig. 5(b) $(\kb T= 0.01\ec)$
is just beginning to be visible in Fig. 5(a) $(\kb T=0.1\ec )$.
It is natural to ask whether even $T=0.01 \ec/\kb$
is the zero temperature limit.
For the higher voltages
and less asymmetric junctions, one has indeed
reached the zero temperature limit,
i.e., the curves do not change appreciably as one goes down in temperature;
however, the lower voltage regions of the more asymmetric junctions
continue to become sharper as one goes down in temperature.
To illustrate this we have plotted in Fig. 6 the region around the first step
in the most asymmetric junction of Figs. 4 and 5
$(\Cl = 0.01 C, \Rl = 0.01 R)$ for four temperatures.
The $T\rightarrow 0$
curves were computed analytically using the two state approximation
of Sec. III. C.  For all of the curves shown,
$I$, $S$, $S/2eI$, and $dI/dV$,
there is significant enhancement of the
structure for temperatures below one hundredth the charging energy.
Both the current and the noise become sharper as one goes down in
temperature, showing that the increased structure in the noise ratio is
a combination of effects in the noise and current.

\subhead{\bf D. Discussion of the noise ratio}

While the above shows that the noise provides
new information about the transport in these systems,
it does not give us a simple interpretation of what
this information means.
In this section we use the analytic results of Sec. III.
to understand our numerical results.
Three limits are discussed: the
thermally activated regime before the first step,
the region just after the first step, where the two-state
approximation is valid, and the large voltage limit,
where the current and noise-voltage characteristics become linear.
We also test an approximate formula for the noise
which interpolates between the high and low voltage limits.
This formula relates the zero frequency noise to the
variance in the number of electrons in the middle region.
All these cases except for the very high voltage limit
are presented in Fig. 7.
The asymptotic large voltage limit is illustrated in Figs. 4 and 5.

The first region we consider is the thermally activated
regime.  In Sec. III. B. we showed that in this region
$S/2eI$ is unity because the current flows in one direction
via rare uncorrelated events.
In Figs. 4 and 5 we do indeed find that the noise ratio
is unity in the low voltage regime below the first step
in the I-V characteristic.
This is shown more explicitly in Fig. 7, where
we replot the solid curves of Figs. 4(b) and 5(b) along
with several analytic approximations, shown as solid dots.
In Fig. 7(a) the thermally activated regime extends up to
$\Cr V/e = 0.5$ (region I).
Because the capacitance ratios vary in Fig. 7(b),
the extent of the thermally activated regime also varies.
For $\Cr > \Cl$ the thermally activated regime extends
up to $\Cr V/e = 1/2$, while for $\Cl > \Cr$ it extends
up to $\Cr V/e = (1/2)(\Cr /\Cl )$.
Thus, the thermally activated regime for the top curve in this figure
only extends up to $\Cr V/e = 0.5 (0.05/0.95) \approx 0.026$.
We have not put in any solid dots for the thermally
activated regime in Fig. 7(b) to avoid cluttering the graph.
In both of these figures the noise ratio is not shown down
to zero voltage because at any finite temperature
($\kb T = 0.01\ec$ here)
thermal fluctuations cause $S/2eI$ to diverge as $V\rightarrow 0$.

The region just above the thermally activated regime in Fig. 7
is the two state limit.
In this regime only two states have nonzero occupation probability.
If the system is in one state, then the only allowed
transition is to the other state.
This introduces correlations which in turn suppress the noise.
For the processes in the thermally activated regime there
were also two relevant rates:
one slow rate due to thermal activation and a quick
decay rate from the excited state.
Because one of the rates was much longer than the other,
the problem reduced to one with just the slow rate.
Another way to say this is that the
noise in the two state limit, $S=2eI((\Gl )^2 +(\Gr )^2)/(\Gl + \Gr )^2$
(Eq. \(2stB)), reduces to classical shot noise, $S=2eI$, when one of the
rates is much larger than the other.

In Figs. 7(a) and (b) one
of the rates is much smaller than the other near the onset,
and the noise ratio is close to unity.
As the voltage is increased
the smaller rate increases and the noise is suppressed.
In this regime the maximum possible suppression is $1/2$
when $\Gr =\Gl$.
In Fig. 7(a) most of the curves
in the two state regime (region II)
do indeed go down to $1/2$.
The curves then return up as the rate which was originally
smaller now becomes the larger rate.
The exceptions are the $\Rl =0.01R$ and $\Cl =0.01C$ junction,
which has not yet reached the zero temperature limit (see Fig. 6),
and the three upper curves, where the smaller rate does not
reach the larger rate before the next step.
In Fig. 7(b) the same physics holds, but we have not illustrated
the two state regime with solid dots because the place where the
two state regime is valid varies with $\Cl /C$.

It is tempting to say that our two state approximation for the
noise works even after the first step, especially for asymmetric
junctions where the probability is strongly peaked about
either the maximum or minimum allowed state.
While the strictly two state approximation predicts the qualitative
features of the noise-ratio voltage curves for higher order steps,
it is not quantitative.  In particular we can see from Figs. 4 and 5
that the noise ratio does not go down to one half after the first
step.  Our two state approximation predicts that the noise goes
down to $1/2$ for the higher steps as well.
Rather than try to extend the two state approximation by, for example,
including three states, we now turn to the large voltage limit.

In Sec. III. D. we argued that in the large voltage limit the noise
ratio should be given by $S/2eI = (\Rl ^{-2}+\Rr ^{-2})/
(\Rl ^{-1}+ \Rr ^{-1})^2$ (Eq. \(2.30)).
This expression is remarkably similar to the expression we found
in the two state limit for the noise ($\Gamma _\ss{L(R)} \rightarrow
R _\ss{L(R)}^{-1}$); however,
it is important to keep in mind that
the high voltage limit is definitely not a limit where the
two state approximation is valid.  Rather, formulas with this
structure appear in a wide variety of contexts.\refto{Ting,Lesovik,earlier}
We regard them as a ubiquitous but not universal high voltage limit
when only two rates are important.
To check this asymptotic limit the dotted lines in Figs. 4 and 5
(not Fig. 7)
are the ratio in Eq. \(2.30).
Clearly, it is approached at large biases.

This still leaves the intermediate regime where more than two
states are important and one has not reached the asymptotic limit
of large voltages.  From Eqs. \(meanl) [$\Rl \le \Rr$]
and \(meanr) [$\Rr \le \Rl$]
we see that at least for asymmetric junctions
the current provides
a measure of the mean number of electrons in the central region.
In region III of Figs. 7(a) and (b) the solid dots over the
numerical I--V characteristics
show that these expressions are good
approximations for the current even for the less asymmetric junctions.
Since the current provides
a measure of the mean number of electrons in the middle region,
$\la n \ra$,
it is natural to try to express the current fluctuations in terms
of the variance of $n$, $\var$.  In
deriving the high voltage limit, we found that
$S=2eI - 4e^2\var /(RC)$ (Eq. \(2.28)).
The solid dots in region III
in Fig. 7(a) and all the solid dots in Fig. 7(b)
are the noise ratio computed with this approximation
using the numerically determined $\var$.
Although this is not as good an approximation as for
the current, it works remarkably well down to lower voltages.
The reductions in the noise can be regarded as an increase in the
variance, which is to be expected near a step where there are large
fluctuations.  The places where the approximation works poorest,
e.g. $R_\ss{L}/R=0.5$ and 0.75 in Fig. 7(a),
are also the places where the approximation for the current
work worst.  This
last approximation for the noise provides us with the simple
intuitive picture that the noise measures the variance in $n$ while
the current measures the mean.

\head{\bf V. Conclusion}
In this paper we have computed the zero frequency current noise for
the simple coulomb blockade problem consisting of two tunnel junctions
connected in series.
The noise was computed both numerically and analytically
using the same master equation as for the current.
The numerical results are exact within the context of
the master equation.  The analytic results were shown to agree with
the numerical results in four regimes:
the zero voltage limit, the thermally activated regime, the
region where only two charge states are accessible at low temperatures,
and the large voltage limit.
The low temperature limit in some cases is not obtained until
the temperature is very much less than $\ec$, e.g. $\kb T = 10^{-3}\ec$.
For the intermediate voltage regime
we obtained a useful approximation
to the noise which related the noise to
the variance in the number of electrons in the region between
the tunnel junctions.  Thus, while the current measures the mean
number of electrons in this region, the noise measures the variance.

Both the numerical and analytic results showed that the noise contains
information which is not contained in the current.
Thus, by measuring the zero frequency noise we can determine
the five parameters in the model more accurately than one
could with the current alone.  Alternately, one can use the
noise to check the results obtained by measuring the current,
showing possible deficiencies in the underlying rate equation
used to describe transport in this system.

The authors would like to thank L. Glazman for useful discussions.
This work was supported primarily
by the U.S. Office of Naval Research and
while the authors were at
the Institute for Theoretical Physics partially
by NSF Grant No. PHY89-04035.
One of us (P.H.) acknowledges the support of the Danish
Natural Science Research Foundation.

\head{\bf Appendix}
The proof of the fluctuation dissipation theorem in quantum
mechanics is few lines long.  On the other hand, the proof for
most classical systems is much longer.
In this appendix we prove the fluctuation dissipation theorem
in two steps.  We first simplify the expression for the
noise in equilibrium and then simplify the expression for
the linear response conductance.

\undertext{Simplify expression for the noise}:
Since we are interested in the zero frequency noise,
we define a new matrix, $\bdp$,
$$
\dP ij = \int _0^\infty dt \left( \P ij (t) - \ro i \right) .\eqno (2.1)
$$
As can be seen by applying $\bm$ to Eq. \(2.1),
$\dP ij$ is related to the inverse of $\bm$:
$$
(\bm \bdp )_{ij} = \ro i - \delta _{ij} .\eqno (2.2)
$$
The zero frequency noise may be expressed directly in terms $\bdp$.
$$
\eqalign{
S_\ss{AB} &= 2e^2 \,\tr{ \bva \bdp \bvb \brz } \cr
                       &+2e^2\,\tr{ \bvb \bdp \bva \brz }  \cr
            &\pm \fancyd 2e^2\, \tr{[\bn ,\bva ]}
             . \cr}
\eqno (2.3)
$$

In equilibrium the matrices
$\bvl$ and $\bvr$ are proportional to one another.
We thus define a matrix $\bv$ which satisfies
$\bvr = \Rl ^{\sm 1}\bv$,
$\bvl = \Rr ^{\sm 1}\bv$, and
$$
\bv = \frac 1{\Rl^{\sm 1} +\Rr^{\sm 1}}\, [\bn,\bm] .
\eqno (2.4)
$$
Eq. \(2.4) allows us to reduce Eq. \(2.3)
to two different kinds of traces:
$\tr {[\bn ,[\bn ,\bm ]]\brz}$ and $\tr{[\bn ,\bm ]\bdp [\bn ,\bm ]\brz}$.
These traces can be simplified using:
(i) the trace of $\bm$ times any vector is zero (Eq. \(1.8)),
(ii) $\bm \brz$ is zero (Eq. \(1.9)),
and (iii) $\bm \bdp$ is related to the identity matrix via Eq. \(2.2).
The end result is that the noise in equilibrium is
$$
S = -4e^2 \frac {\Rl ^{\sm 1}\Rr ^{\sm 1}}{(\Rl^{\sm 1}+\Rr^{\sm 1})^2}
\, \tr {\bn \bm \bn \brz}. \eqno (2.5)
$$

\undertext{Simplify expression for the linear response conductance}:
In equilibrium the probability function, $\rho ^\ss{(0)}_n$,
has a thermal distribution,
i.e. it is the exponential of $-\beta$ times some energy, $E _n$.
The simplest way to see this is to use detailed balance (Eq. \(1.23))
$$
\frac {\g n{n-1}}{\g {n-1}n} = \frac {\ro {n-1}}{\ro n}
=\frac {e^{-\beta E_{n-1}}}{e^{-\beta E_n}},
\eqno (2.6)
$$
where the energy $E_n$ is
$$
E_n = \frac {e^2n^2}{2C} + e\Vp n .
\eqno (2.7)
$$
The fact that the distribution is thermal and that $E_n$ is
given by Eq. \(2.7) allows one to express the derivative of
$\brz$ with respect to $\Vp$ in terms of the equilibrium
probability:
$$
\dvx {\ro n} = -\beta e \,
\left( n - \tr {\bn \brz}\ro n \right) .
\eqno (2.8)
$$

The strategy now is to express
partial derivatives with
respect to $V$ in terms of partial derivatives with respect to
$\Vp$.  At the end we use Eq. \(2.8) to rewrite partials with respect
to $\Vp$ in terms of equilibrium expectation values.
Because we know that the currents $\il$ and $\ir$ are equal, we
will only consider $\il$.
The quantity which we are computing is the
linear response conductance:
$$
\dv I = -e \,\tr{\dv \bvl \brz} -e \,\tr{\bvl \dv \brz} ,
\eqno (2.9)
$$
where this and all other partial derivatives are evaluated at $V=0$.
The derivative of $I$ with respect to $\Vp$ at $V=0$ is zero
because $I$ at $V=0$ is zero.
$$
\dvx I = -e \tr{\dvx \bvl \brz} -e \tr{\bvl \dvx \brz}=0
\eqno (2.10)
$$
Because the
$\Gamma ^\ss{L}$'s are functions of $\Cr V + C\Vp$,
derivatives with respect to $V$ can be related to
ones with respect to $\Vp$.
$$
\dv \bvl = \frac {\Cr}{C} \dvx \bvl
\eqno (2.11)
$$
Using Eqs. \(2.9) - \(2.11), one partial derivative
with respect to $V$ can be eliminated:
$$
\dv I = -e\,\tr {\bvl \left(\dv \brz - \frac {\Cr}{C} \dvx \brz \right) } .
\eqno (2.12)
$$

In order to eliminate the other derivative with respect to
$V$, we note that $\bm \brz$ is zero for all voltages and hence
$$
\eqalignno{
\dv M \brz + \bm \dv \brz &=0 &(2.13a) \cr
\dvx M \brz + \bm \dvx \brz &=0. &(2.13b) \cr }
$$
As in Eq. \(2.11), we can relate derivatives with
respect to $V$ to derivatives with respect to $\Vp$:
$$
\dv \bm \, =\, \frac {\Rl ^{\sm 1} \Cr - \Rr ^{\sm 1} \Cl}
            {C (\Rl ^{\sm 1} + \Rr ^{\sm 1})}   \,\dvx \bm .
\eqno (2.14)
$$
Using Eqs. \(2.13a) -\(2.14), the linear response conductance is
$$
\dv I\, =\, e\,\frac {\Rl ^{\sm 1}\Rr ^{\sm 1}}{(\Rl ^{\sm 1} + \Rr ^{\sm
1})^2}
\,\, \tr{ \bn \bm \dvx \brz} .
\eqno (2.15)
$$
As our final simplification of $dI/dV$, we use Eq. \(2.8) to covert
the partial with respect to $\Vp$ to traces involving $\brz$.
$$
\dv I=-\beta e^2\,\frac{\Rl^{\sm 1}\Rr^{\sm 1}}{(\Rl ^{\sm 1} + \Rr ^{\sm
1})^2}
\,\,\tr{\bn \bm \bn \brz}
\eqno (2.16)
$$
Dividing Eq. \(2.5) by Eq. \(2.16) we obtain the
the fluctuation dissipation theorem, Eq. \(FDT).

\references
\item{\dag}  Permanent address: Department of Electronics and Electrical
Engineering, University of Glasgow, Glasgow G12 8QQ, UK.

\item{\ddag} Permanent address: Department of Physics, University of Florida,
Gainsville, FL 32611.

\refis{Johnson} J. B. Johnson, {\it Phys. Rev.} {\bf 29}, 367 (1927).

\refis{Nyquist} H. Nyquist, {\it Phys. Rev.} {\bf 32}, 229 (1928).

\refis{classicalinbook} For a review see
K. M. van Vliet and J. R. Fassett, {\it Fluctuation
Phenomena in Solids} (ed. R. E. Burgess), p. 267, Academic Press,
New York, 1965.

\refis{LandNoise} For a discussion in the context of tunneling see
R. Landauer, {\it Jour. Appl. Phys.} {\bf 33}, 2209 (1962).

\refis{earlier} J. H. Davies, P. Hyldgaard, S. Hershfield,
and J. W. Wilkins, to be published.

\refis{Clarke} A. N. Cleland, J. M. Schmidt, and J. Clarke,
{\it Physica B} {\bf 165\&166}, 979 (1990).

\refis{Mooij} L. J. Geerligs, et. al., {\it Phys. Rev. Lett.} {\bf 64},
2691 (1990).

\refis{Likharev} K. K. Likharev, {\it IBM J. Res. Dev.} {\bf 32}, 144 (1988).

\refis{oldCB1} H. R. Zeller and I. Giaever, {\it Phys. Rev.} {\bf 181},
789 (1969).

\refis{oldCB2} J. Lambe and R. C. Jaklevic, {\it Phys. Rev. Lett.} {\bf 22},
1371 (1969).

\refis{Fulton} D. A. Fulton and G. J. Dolan, {\it Phys. Rev. Lett.}
{\bf 59}, 109 (1987).

\refis{manyStair} J. B. Barner and S. T. Ruggiero, {\it Phys. Rev. Lett.}
{\bf 59}, 807 (1987).

\refis{fewStair} L. S. Kuz'min and K. K. Likharev, {\it Pis'ma Zh. Eksp. Teor.
Fiz.} {\bf 45}, 389 (1987) [{\it JETP Lett.} {\bf 45}, 495 (1987)].

\refis{oneStair} P. J. M. van Bentum, T. T. M. Smokers, and H. van Kempen,
{\it Phys. Rev. Lett.} {\bf 60}, 2543 (1988).

\refis{STM} R. Wilkins, E. Ben-Jacob, and R. C. Jaklevic, {\it Phys. Rev.
Lett.} {\bf 63}, 801 (1989).

\refis{Gregory} S. Gregory, {\it Phys. Rev. Lett.} {\bf 64}, 689 (1990).

\refis{Webb} V. Chandrasekhar, Z. Ovadyahu, and R. A. Webb, {\it Phys. Rev.
Lett.} {\bf 67}, 2862 (1991).

\refis{originalcalc} O. Kulik and R. I. Shekhter {\it Zh. Eksp. Teor. Fiz.}
{\bf 68}, 623 (1975) [{\it Sov. Phys. JETP} {\bf 41}, 308 (1975)].

\refis{numericalcalc} K. Mullen, E. Ben-Jacob, R. C. Jaklevic, and
Z. Schuss, {\it Phys. Rev. B} {\bf 37}, 98 (1988).

\refis{analyticcalc} M. Amman, R. Wilkins, E. Ben-Jacob, P. D. Maker,
and R. C. Jaklevic, {\it Phys. Rev. B} {\bf 43}, 1146 (1991).

\refis{twostate} J-C. Wan, K. A. McGreer, L. I. Glazman, A. M. Goldman,
and R. I. Shekhter, {\it Phys. Rev. B} {\bf 43}, 9381 (1991).

\refis{Tsui} Y. P. Li, A. Zaslavsky, D. C. Tsui, M. Santos,
and M. Shayegan, {\it Phys. Rev. B} {\bf 41}, 8388 (1990).

\refis{Lesovik} G. B. Lesovik, {\it Pis'ma. Zh. Eksp. Teor. Fiz.} {\bf 49},
513 (1989) [{\it JETP Letters} {\bf 49}, 592 (1989)].

\refis{Ting} Chen, L. Y. and Ting, C. S., {\it Phys. Rev. B} {\bf 43},
534 (1991).

\refis{Rogers} C. T. Rogers and R. A. Buhrman, {\it Phys. Rev. Lett.}
{\bf 53}, 1272 (1984).

\refis{noiseexp} G. Zimmerli, T. M. Eiles, H. D. Jensen, R. L. Kautz,
and J. M. Martinis, {\it Bull. Am. Phys. Soc.} {\bf 37}, 764 (1992).

\refis{noisecalc} A. N. Korothov, D. V. Averin, K. K. Likharev, and
S. A. Vasenko, preprint.

\refis{ournoisecalc} J. W. Wilkins, S. Hershfield, J. H. Davies, P.
Hyldgaard, and C. Stanton, in {\it Proceedings of the Nobel Jubilee Symposium
on Low Dimensional Properties of Solids,} (Gothenburg, Sweden,
December 4-7, 1991); {\it Bull. Am. Phys. Soc.} {\bf 37}, 661 (1992).

\endreferences

\head{\bf Figure Captions}

\item{Fig. 1} Schematic of experimental geometry and rates.
(a) The capacitances and resistances
of the left and right junctions are $(\Cl ,\Cr)$ and $(\Rl ,\Rr)$,
respectively.  The voltage in the left and right leads are fixed
to $V$ and 0, respectively.
The voltage in the central region, $V_m$, fluctuates depending on
the number of electrons in the central region, $n$.
(b) At zero temperature the rates for tunneling
are linear in $n$.  For a positive bias, $V>0$, only the rates
for tunneling onto the central region from the right lead,
$\gr n{n+1}$,
and the rates for tunneling off the central region to the
left lead, $\gl n{n-1}$ are nonzero.
The points where these rates vanish, $\Nr$ and $\Nl$,
determine the maximum and minimum charge states,
$N_{max}$ and $N_{min}$.

\item{Fig. 2} The equilibrium noise, $S$, and conductance,
$G = dI/dV$, for two tunnel junctions connected in series.
At temperatures far below the charging energy
of the tunnel junctions, $\ec = e^2/2C$, both the conductance (solid curve)
and the zero-frequency
noise (dotted curve) are suppressed by the coulomb blockade.
At temperatures above the charging energy the conductance
is roughly linear in temperature and the noise approaches
a constant value.
In accordance with the fluctuation dissipation theorem
the ratio of the noise to the conductance remains fixed
at $4\kb T$ (see dashed curve).

\item{Fig. 3} The nonequilibrium noise and current.
At low temperatures ($\kb T=0.01\ec$)
for an asymmetric junction ($\Rl/R = 0.01$ and $\Cl/C = 0.01$)
both the average current, $I$,
and the noise, $S$, show the characteristic step structure of the coulomb
staircase. Indeed, $S$ is roughly equal to the standard
shot noise result of $2eI$.
With an offset voltage, $\Vp$,
the current-voltage and noise-voltage
characteristics are shifted,
but other qualitative features of the curves remain
the same.  Henceforth we set $\Vp =0$.

\item{Fig. 4} The dependence of the
current and noise ratio ($S/2eI$) on the resistances of the
junctions at (a) $\kb T = 0.1\ec$ and (b) $\kb T = 0.01 \ec$.
We go from good steps ($\Cl/C=0.01$,$\Rl/R=0.01$)
to no steps ($\Cl/C=0.01$, $\Rl/R=0.99$)
by varying the resistance ratio, $\Rl/\Rr$, keeping
the capacitances fixed at $\Cl/C=0.01$ and $\Cr/C=0.99$.
When the difference between $S$ and $I$ illustrated in Fig. 3 is
plotted as the noise ratio, $S/2eI$, a rich variety of
structure is revealed.
This structure becomes more pronounced
as one goes down in temperature.
The noise ratio shows structure even when
the current shows little structure,
illustrating how the noise can provide new information about the
parameters in the model.
The dotted lines are the asymptotic value of the noise ratio
at large voltages (Eq. \(2.30)).
Since the noise remains finite as the current goes to zero,
the noise ratio diverges as $V\rightarrow 0$.
We have thus only plotted it for $\Cr V/e > 0.2$.
(This is the reason $S/2eI >1$ at low voltages in (a).)

\item{Fig. 5} The current and noise ratio as a function of
the capacitance ratio at (a) $\kb T = 0.1\ec$ and (b) $\kb T = 0.01 \ec$.
$(\Rl /R = 0.01)$
Varying the capacitance instead of the resistance has a
different effect.  All the noise ratio curves have a similar
structure, indicating that
the noise is more sensitive to the the resistance ratio than
capacitance ratio.  This should not be too surprising because the
asymptotic limit of noise ratio (dotted lines)
is determined by the resistances of
the junctions (see Eq. \(2.30)).
Although the abscissas of Fig. 4 and this figure are the same,
the voltage ranges are different because the capacitances vary here.

\item{Fig. 6} The temperature dependence
for the most asymmetric junctions shown in Figs. 4 and 5
($\Rl /R = 0.01$, $\Cl /C = 0.01$).
The current ($I$), the noise ($S$), the noise ratio ($S/2eI$),
and the differential
conductance ($G=dI/dV$) all show significant temperature dependence below
$\kb T = 0.01\ec$.
The spiky
structure in the noise ratio is due to a combined effect of $I$
and $S$ having sharper structure as one goes to lower temperatures.
The zero temperature curves are obtained from the
analytic expressions for the noise in the two state limit
(Eqs. \(2stA) and \(2stB)).

\item{Fig. 7} Approximations for the current and the noise.
The solid lines are the numerical results of Figs. 4(b) and 5(b),
and the dots are the approximations.
(a) In the thermally activated regime labeled by I
the noise ratio is unity.  In the two state regime
labeled by II the current and noise
are given by Eqs. \(2stA) and \(2stB).
Both this and the thermally activated regime are exact at low temperatures.
In the higher voltage regime, region III, we have used
Eqs. \(meanl) [$\Rl \le \Rr$] -\(meanr) [$\Rr \le \Rl$]  and Eq. \(2.28)
which relate the current and noise to the mean and variance in the
number of electrons in the region between the two tunnel junctions.
Although these approximations are not exact, they clearly get the important
qualitative features of the current and noise.
(b) With this abscissa the
boundaries of the thermally activated and two-state regimes vary
as one changes the capacitances.
Thus, although the thermally activated and two state regimes are also
exact here, we have only shown the large voltage approximation to
avoid cluttering the graph.
\endpaper
\end